\documentclass{article}
\usepackage{graphicx}
\usepackage{amssymb}
\usepackage{amsfonts}
\usepackage{amsmath}

\begin{document}

\title{Coupled nonlinear oscillators: metamorphoses of amplitude profiles
for the approximate effective equation -- the case of $1:3$ resonance}
\author{Jan Kyziol$^{1)}$, Andrzej Okninski$^{2)}$ \\
Department of Mechatronics and Mechanical Engineering$^{1)}$, \\
Physics Division, Department of Management and Computer Modelling$^{2)}$, \\
Politechnika Swietokrzyska, Al. 1000-lecia PP 7, \\
25-314 Kielce, Poland}
\maketitle

\begin{abstract}
We study dynamics of two coupled periodically driven oscillators. An
important example of such a system is a dynamic vibration absorber which
consists of a small mass attached to the primary vibrating system of a large
mass.

Periodic solutions of the approximate effective equation (derived in our
earlier papers) are determined within the Krylov-Bogoliubov-Mitropolsky
approach to compute the amplitude profiles $A\left( \Omega \right) $. In the
present paper we investigate metamorphoses of the function $A\left( \Omega
\right) $ induced by changes of the control parameters in the case of $1:3$
resonances.
\end{abstract}

\section{Introduction}

In the present paper we analyse two coupled oscillators, one of which is
driven by an external periodic force. An important example of such a system
is a dynamic vibration absorber which consists of a mass $m_{2}$ , attached
to the primary vibrating system of mass $m_{1}$ \cite%
{DenHartog1985,Oueini1999}. Equations describing dynamics of this system are
of form: 
\begin{equation}
\left. 
\begin{array}{l}
m_{1}\ddot{x}_{1}-V_{1}\left( \dot{x}_{1}\right) -R_{1}\left( x_{1}\right)
+V_{2}\left( \dot{x}_{2}-\dot{x}_{1}\right) +R_{2}\left( x_{2}-x_{1}\right)
=f\cos \left( \omega t\right) \\ 
m_{2}\ddot{x}_{2}-V_{2}\left( \dot{x}_{2}-\dot{x}_{1}\right) -R_{2}\left(
x_{2}-x_{1}\right) =0%
\end{array}%
\right\}  \label{model1}
\end{equation}%
where $V_{1}$, $R_{1}$ and $V_{2}$, $R_{2}$ represent (nonlinear) force of
internal friction and (nonlinear) elastic restoring force for mass $m_{1}$
and mass $m_{2}$, respectively. In the present paper we shall consider a
simplified model:%
\begin{equation}
R_{1}\left( x_{1}\right) =-\alpha _{1}x_{1},\ V_{1}\left( \dot{x}_{1}\right)
=-\nu _{1}\dot{x}_{1}.  \label{simplified}
\end{equation}

Dynamics of coupled periodically driven oscillators is very complicated, see 
\cite%
{Szemplinska1990,Awrejcewicz1991,Kozlowski1995,Janicki1995,Kuznetsov2009}
and references therein. We simplified the problem described by equations (%
\ref{model1}), (\ref{simplified}) deriving the exact fourth-order nonlinear
equation for internal motion as well as approximate second-order effective
equation in \cite{Okninski2006}.

Moreover, applying the Krylov-Bogoliubov-Mitropolsky method to these
equations we have computed the corresponding nonlinear resonances in the
effective equation (cf. \cite{Okninski2006} and \cite{Kyziol2007} for the
cases of $1:1$ and $1:3$\ resonances, respectively). Dependence of the
amplitude $A$ of nonlinear resonances on the frequency $\omega $ is
significantly more complicated than in the case of Duffing oscillator and
this leads to new nonlinear phenomena. In a recent paper we investigated
metamorphoses of the function $A\left( \omega \right) $ induced by changes
of the control parameters in the case of $1:1$ resonance \cite{Kyziol2011}.
In the present paper we continue this approach studying metamorphoses of $%
A\left( \omega \right) $ for $1:3$ resonance.

In the next Section the exact 4th-order equation for the internal motion and
approximate 2nd-order effective equations in non-dimensional form are
described. In Section 3 amplitude profiles for $1:3$ resonances\ are
determined within the Krylov-Bogoliubov-Mitropolsky approach for the
approximate 2nd-order effective equation (and for the Duffing equation which
follows from the effective equation if some parameters are put equal zero).
In Section 4 theory of algebraic curves is used to compute singular points
of effective equation amplitude profiles -- metamorphoses of amplitude
profiles occur in neighbourhoods of such points. In Section 5 examples of
analytical and numerical computations are presented for the Duffing
equation. Our results are summarized and perspectives of further studies are
described in the last Section.

\section{Exact equation for internal motion and its approximations}

In new variables, $x\equiv x_{1}$, $y\equiv x_{2}-x_{1}$, equations (\ref%
{model1}), (\ref{simplified}) can be written as:

\begin{equation}
\left. 
\begin{array}{l}
m\ddot{x}+\nu \dot{x}+\alpha x+V_{e}\left( \dot{y}\right) +R_{e}\left(
y\right) =f\cos \left( \omega t\right) \\ 
m_{e}\left( \ddot{x}+\ddot{y}\right) -V_{e}\left( \dot{y}\right)
-R_{e}\left( y\right) =0%
\end{array}%
\right\} ,  \label{model2}
\end{equation}%
where $m\equiv m_{1}$, $m_{e}\equiv m_{2}$, $\nu \equiv \nu _{1}$, $\alpha
\equiv \alpha _{1}$, $V_{e}\equiv V_{2}$, $R_{e}\equiv R_{2}$. It is
possible to simplify the problem eliminating the variable $x$ in (\ref%
{model2}) to obtain the exact fourth-order equation for the variable $y$
only -- describing relative motion of the mass $m_{e}$ \cite{Okninski2006}.

In the present work we assume:%
\begin{equation}
R_{e}\left( y\right) =\alpha _{e}y-\gamma _{e}y^{3},\quad V_{e}\left( \dot{y}%
\right) =-\nu _{e}\dot{y}.  \label{ReVe}
\end{equation}%
The exact equation for relative motion reads:%
\begin{equation}
\begin{array}{l}
\hat{L}\,\left( \mu \tfrac{d^{2}y}{dt^{2}}+\nu _{e}\tfrac{dy}{dt}-\alpha
_{e}y+\gamma _{e}y^{3}\right) +\lambda m_{e}\left( \nu \tfrac{d}{dt}+\alpha
\right) \tfrac{d^{2}y}{dt^{2}}=F\cos \left( \omega t\right) , \\ 
\hat{L}\equiv M\frac{d^{2}}{dt^{2}}+\nu \frac{d}{dt}+\alpha ,%
\end{array}
\label{4th}
\end{equation}%
where $F=m_{e}\omega ^{2}f$, $\mu =mm_{e}/M$ and $\lambda =m_{e}/M$ is a
nondimensional parameter \cite{Okninski2006}.

Eqn. (\ref{4th}) can be written in the following nondimensional form \cite%
{Kyziol2011}:%
\begin{equation}
\begin{array}{l}
\mathcal{\hat{L}}\,\left( \tfrac{d^{2}z}{d\tau ^{2}}+h\tfrac{dz}{d\tau }%
-z+z^{3}\right) +\kappa \left( H\tfrac{d}{d\tau }+a\right) \tfrac{d^{2}z}{%
d\tau ^{2}}=G\tfrac{\kappa }{\kappa +1}\Omega ^{2}\cos \left( \Omega \tau
\right) , \\ 
\mathcal{\hat{L}}\equiv \tfrac{d^{2}}{d\tau ^{2}}+H\tfrac{d}{d\tau }+a,%
\end{array}
\label{4th-nondim}
\end{equation}%
where nondimensional time $\tau $ and nondimensional displacement $z$ of the
mass $m_{e}$ are defined as:

\begin{equation}
\tau =t\bar{\omega},\ z=y\sqrt{\tfrac{\gamma _{e}}{\alpha _{e}}},\qquad
\left( \bar{\omega}=\sqrt{\tfrac{\alpha _{e}}{\mu }}\right)  \label{Ndim1}
\end{equation}%
while nondimensional constants are given by:%
\begin{equation}
h=\tfrac{\nu _{e}}{\mu \bar{\omega}},\ H=\tfrac{\nu }{M\bar{\omega}},\
\Omega =\tfrac{\omega }{\bar{\omega}},\ G=\tfrac{1}{\alpha _{e}}\sqrt{\tfrac{%
\gamma _{e}}{\alpha _{e}}}f,\ \kappa =\tfrac{m_{e}}{m},\ a=\tfrac{\alpha \mu 
}{\alpha _{e}M}.  \label{Ndim2}
\end{equation}

We shall consider hierarchy of approximate equations arising from (\ref%
{4th-nondim}) \cite{Kyziol2011}. For small enough values of the parameters $%
\kappa ,\ H,\ a$ we can reject the second term on the left in (\ref%
{4th-nondim}) obtaining the approximate equation which can be integrated
partly to yield the effective equation:

\begin{equation}
\frac{d^{2}z}{d\tau ^{2}}+h\frac{dz}{d\tau }-z+z^{3}=-\gamma \tfrac{\Omega
^{2}}{\sqrt{\left( \Omega ^{2}-a\right) ^{2}+H^{2}\Omega ^{2}}}\cos \left(
\Omega \tau +\delta \right) ,\qquad \left( \gamma \equiv G\tfrac{\kappa }{%
\kappa +1}\right)  \label{effective}
\end{equation}%
where transient states have been omitted and $\tan \delta =\tfrac{\Omega H}{%
\Omega ^{2}-a}$. And, finally, for $H=0,\ a=0$ we get the Duffing equation:%
\begin{equation}
\frac{d^{2}z}{d\tau ^{2}}+h\frac{dz}{d\tau }-z+z^{3}=-\gamma \cos \left(
\Omega \tau +\delta \right) .  \label{Duffing}
\end{equation}

\section{Perturbation analysis of the $1:3$ resonance}

The $1:3$ resonance, a solution of the effective equation (\ref{effective})
of form $z=A\cos \left( 3\Omega \tau +\varphi \right) $, can be seen in the
bifurcation diagram computed for the effective equation -- see Fig. $4.1$ in 
\cite{Okninski2006}, $\omega < 3.1$. We apply the
Krylov-Bogoliubov-Mitropolsky (KBM) perturbation approach \cite{Nayfeh1981,
Awrejcewicz2006} to Eqn. (\ref{effective}), working in the spirit of \cite%
{Janicki1995}, to determine the corresponding amplitude profile, i.e.
dependence of the amplitude $A$ on frequency $\Omega $.

To study subharmonic resonance $1:3$ we cast equation (\ref{effective}) into
form:

\begin{equation}
\tfrac{d^{2}z}{d\tau ^{2}}+\Theta ^{2}z+\varepsilon \left( \left(
-a_{0}-\Theta _{0}^{2}+a_{0}z^{2}\right) z+h_{0}\tfrac{dz}{d\tau }\right) =%
\tfrac{-\gamma \Omega ^{2}\cos \left( \Omega \tau +\delta \right) }{\sqrt{%
\left( \Omega ^{2}-a\right) ^{2}+H^{2}\Omega ^{2}}}  \label{prep1a}
\end{equation}%
with%
\begin{equation}
\varepsilon \Theta _{0}^{2}=\Theta ^{2},\ \varepsilon a_{0}=1,\ \varepsilon
h_{0}=h,  \label{prep1b}
\end{equation}%
where we assumed that the external force is of order $\varepsilon ^{0}$
rather than $\varepsilon ^{1}$ (see \cite{Bogoliubov1961} for discussion).

We substitute $z\left( \tau \right) =u\left( \tau \right) +u_{0}\left( \tau
\right) $ into (\ref{prep1a}) to remove the external forcing term on the
right-hand side. We thus get:%
\begin{equation}
\begin{array}{l}
\frac{d^{2}u}{d\tau ^{2}}+\Theta ^{2}u+\frac{d^{2}u_{0}}{d\tau ^{2}}+\Theta
^{2}u_{0}+\varepsilon g\left( u,u_{0}\right) =\tfrac{-\gamma \Omega ^{2}\cos
\left( \Omega \tau +\delta \right) }{\sqrt{\left( \Omega ^{2}-a\right)
^{2}+H^{2}\Omega ^{2}}}, \\ 
g\left( u,u_{0}\right) =h_{0}\frac{d\left( u+u_{0}\right) }{d\tau }+\left(
-a_{0}-\Theta _{0}^{2}\right) \left( u+u_{0}\right) +a_{0}\left(
u+u_{0}\right) ^{3}.%
\end{array}
\label{prep1c}
\end{equation}

Now we put $u_{0}\left( \tau \right) =C\cos \left( \Omega \tau +\delta
\right) $ into (\ref{prep1c}). It follows that for $C=\tfrac{-\gamma \Omega
^{2}}{\sqrt{\left( \Omega ^{2}-a\right) ^{2}+H^{2}\Omega ^{2}}}\tfrac{1}{%
\Theta ^{2}-\Omega ^{2}}$ two terms on the left-hand side, $\frac{d^{2}u_{0}%
}{d\tau ^{2}}+\Theta ^{2}u_{0}$, and the external forcing term on the
right-hand side of (\ref{prep1c}) cancel out to yield:%
\begin{equation}
\tfrac{d^{2}u}{d\tau ^{2}}+\Theta ^{2}u+\varepsilon g\left( u,u_{0}\right)
=0.  \label{prep1d}
\end{equation}

We shall now determine approximate form of $\Theta ^{2}$ following procedure
described in \cite{Janicki1995}. Neglecting in (\ref{prep1a}) the damping
term $h_{0}\tfrac{dz}{d\tau }$ and external forcing we get:%
\begin{equation}
\tfrac{d^{2}z}{d\tau ^{2}}-z+z^{3}=0.  \label{appr}
\end{equation}%
Substituting in (\ref{appr}) $z\left( \tau \right) =A\cos \left( \Theta \tau
\right) $, applying identity $\cos ^{3}\left( \Theta \tau \right) =\frac{3}{4%
}\cos \left( \Theta \tau \right) +\frac{1}{4}\cos \left( 3\Theta \tau
\right) $, and rejecting term proportional to $\cos \left( 3\Theta \tau
\right) $ we get finally the approximate expression $\Theta ^{2}=\frac{3}{4}%
A^{2}-1$.

We have thus written the effective equation (\ref{effective}) in form (\ref%
{prep1d}) with $g\left( u,u_{0}\right) $ defined in (\ref{prep1c}) and:%
\begin{equation}
u_{0}\left( \tau \right) =C\cos \left( \Omega \tau +\delta \right) ,\quad C=%
\tfrac{-\gamma \Omega ^{2}}{\sqrt{\left( \Omega ^{2}-a\right)
^{2}+H^{2}\Omega ^{2}}}\tfrac{1}{\Theta ^{2}-\Omega ^{2}},\quad \Theta ^{2}=%
\tfrac{3}{4}A^{2}-1.  \label{prep1e}
\end{equation}

Since we are looking for $1:3$ resonances we have to consider frequencies $%
\Omega $ close to $3\Theta $. We thus put in (\ref{prep1d}) $\Theta
^{2}=\left( \tfrac{\Omega }{3}\right)^{2}+\varepsilon \sigma $ with $\sigma $
of order $\varepsilon ^{0}$, obtaining finally 
\begin{equation}
\tfrac{d^{2}u}{d\tau ^{2}}+\left( \tfrac{\Omega }{3}\right)
^{2}u+\varepsilon \left( \sigma u+g\left( u,u_{0}\right) \right) =0.
\label{final}
\end{equation}

We assume the following form of the solution: 
\begin{equation}
u=A\cos \left( \tfrac{\Omega }{3}\tau +\varphi \right) +\varepsilon
u_{1}\left( A,\varphi ,\tau \right) +\ldots \ .  \label{sol1}
\end{equation}

Substituting (\ref{sol1}) into (\ref{final}), eliminating secular terms and
demanding that $\frac{dA}{d\tau }=0$, $\frac{d\varphi }{d\tau }=0$ to find
stationary states we get finally \cite{Kyziol2007}:%
\begin{equation}
\begin{array}{l}
\left( h\tfrac{\Omega }{3}\right) ^{2}+\left( \frac{3}{4}A^{2}+\frac{3}{2}%
C^{2}-\frac{1}{9}\Omega ^{2}-1\right) ^{2}=\left( \frac{3}{4}AC\right) ^{2},
\\ 
\tan \left( 3\varphi -\delta \right) =\dfrac{-h\Omega }{3\left( \frac{3}{4}%
A^{2}+\frac{3}{2}C^{2}-\frac{1}{9}\Omega ^{2}-1\right) },%
\end{array}
\label{AmplEff}
\end{equation}%
with $C$ given by (\ref{prep1e}). If we put $H=0$, $a=0$ then we get
implicit equation for the amplitude profile for the Duffing equation:%
\begin{equation}
\left( h\tfrac{\Omega }{3}\right) ^{2}+\left( \tfrac{3}{4}A^{2}+\tfrac{3}{2}%
C^{2}-\tfrac{1}{9}\Omega ^{2}-1\right) ^{2}=\left( \tfrac{3}{4}AC\right)
^{2},\ C=\dfrac{-\gamma }{\left( \frac{3}{4}A^{2}-\Omega ^{2}-1\right) }.
\label{AmplDuff}
\end{equation}

\section{Metamorphoses of the amplitude profiles for the $1:3$ resonance}

Equations (\ref{AmplEff}), (\ref{AmplDuff}) define the corresponding
amplitude profiles implicitly. Such amplitude profiles can be classified as
planar algebraic curves, see \cite{Wall2004} for a general theory. Let $%
L\left( X,Y;\lambda \right) =0$ defines such a curve where $\lambda $ is a
parameter. A singular point $\left( X_{0},Y_{0}\right) $ of the algebraic
curve obeys conditions:%
\begin{equation}
L\left( X,Y;\lambda \right) =0,\qquad \frac{\partial L\left( X,Y;\lambda
\right) }{\partial X}=0,\qquad \frac{\partial L\left( X,Y;\lambda \right) }{%
\partial Y}=0.  \label{Singular}
\end{equation}%
Assume that a solution $\left( X_{0},Y_{0}\right) $ of Eqns. (\ref{Singular}%
) exists for $\lambda =\lambda _{0}$ and there are no other solutions in
some neighbourhood of $\lambda _{0}$. Let $\lambda <\lambda _{0}$, then the
curve $L\left( X,Y;\lambda \right) =0$ for growing values of $\lambda $
changes its form at $\lambda =\lambda _{0}$ and, again, for $\lambda
>\lambda _{0}$. We shall refer to such changes as metamorphoses (cf. \cite%
{Kyziol2011} for metamorphoses of amplitude profiles in the case of $1:1$
resonance in the effective equation).

In the case of the effective equation the amplitude profile of $1:3$
resonance is given by Eqn. (\ref{AmplEff}) or, in new variables $X\equiv
\Omega ^{2}$, $Y\equiv A^{2}$, by the equation $L\left( X,Y;a,\gamma
,h,H\right) =0$ where
\begin{equation}
\begin{array}{l}
L\left( X,Y;a,\gamma ,h,H\right) =U^{4}\left( \tfrac{1}{9}%
h^{2}X+U^{2}\right) \left( \left( X-a\right) ^{2}+H^{2}X\right) ^{2}+ \\ 
+3\gamma ^{2}U^{2}X^{2}\left( \tfrac{9}{16}Y-\tfrac{1}{9}X-1\right) \left(
\left( X-a\right) ^{2}+H^{2}X\right) +\tfrac{9}{4}\gamma ^{4}X^{4}, \\ 
U\equiv \frac{3}{4}Y-X-1.%
\end{array}
\label{LEff}
\end{equation}

Equations for singular points of the amplitude profile for the $1:3$\
resonance of the effective equation are given by (\ref{Singular}), (\ref%
{LEff}). To find solutions of these equations we solve the following cubic
equation:%
\begin{equation}
\begin{array}{l}
c_{3}U^{3}+c_{2}U^{2}+c_{1}U+c_{0}=0, \\ 
c_{3}=567B_{1}^{2} \\ 
c_{2}=-18B_{1}\left( 28X^{3}+\left( 87q-81\right) X^{2}+\left(
146r-54q\right) X-27r\right) \\ 
c_{1}=-4B_{2}X\left( 967X^{3}+639X^{2}+\left( 162-967r+423q\right)
X+207r+81q\right) \\ 
c_{0}=32B_{2}^{2}X\left( 55X^{2}+9\right) \\ 
B_{1}=3X^{2}+qX-r,\ B_{2}=X^{2}+qX+r,\ q=H^{2}-2a,\ r=a^{2},%
\end{array}
\label{solU}
\end{equation}%
for arbitrary $X>0,$ $a\geq 0,$ $H\geq 0$ and compute variables $p,\ s$: 
\begin{eqnarray}
p &=&-\tfrac{2}{9}\tfrac{\left( 9U+8X\right) \left( \left(
3X^{3}+2X^{2}q+\left( -9q+r\right) X-18r\right)
U-16X^{3}-16X^{2}q-16Xr\right) }{X\left( \left( 81X^{2}+27qX-27r\right)
U-46X^{3}+\left( -18-46q\right) X^{2}+\left( -46r-18q\right) X-18r\right) },
\notag \\
s &=&-\tfrac{4}{9}\tfrac{U^{2}\left( X^{2}+qX+r\right) \left(
128X^{2}+162pX+360XU+243U^{2}\right) }{X^{2}\left( 46X+81U-18\right) }.
\label{solps}
\end{eqnarray}

Finally, we find $Y,\ h$ and $\gamma $ from definitions:%
\begin{equation}
U=\tfrac{3}{4}Y-X-1,\ p=\tfrac{1}{9}h^{2},\ s=\tfrac{3}{2}\gamma ^{2},
\label{definitions}
\end{equation}
and physically acceptable solutions must fulfill conditions: $Y>0,\ h>0,\
\gamma >0$.

We can obtain the case of the Duffing equation putting $a=H=0$ in the above
formulae. We thus get:%
\begin{equation}
L\left( X,Y;\gamma ,h\right) =U^{4}\left( \tfrac{1}{9}h^{2}X+U^{2}\right)
+3\gamma ^{2}U^{2}\left( \tfrac{9}{16}Y-\tfrac{1}{9}X-1\right) +\tfrac{9}{4}%
\gamma ^{4},  \label{LDuff}
\end{equation}%
and
\begin{equation}
\begin{array}{l}
5103U^{3}+\left( -1512X+4374\right) U^{2}+\left( -3868X^{2}-2556X-648\right)
U \\ 
+1760X^{3}+288X=0%
\end{array}
\label{sol2a}
\end{equation}%
where $X$ is arbitrary, $U=\tfrac{3}{4}Y-X-1$ and%
\begin{equation}
h^{2}=2\tfrac{24UX+27U^{2}-128X-144U}{46X-81U+18},\quad \gamma ^{2}=\tfrac{8%
}{27}U^{2}\tfrac{72UX+243U^{2}-128X^{2}}{46X-81U+18}.  \label{sol2b}
\end{equation}

\section{Analytical and numerical computations: the Duffing equation}

We shall find a metamorphosis of the bifurcation diagram for the Duffing
equation (\ref{Duffing}). To this end we have to compute a singular point of
the amplitude profile (\ref{AmplDuff}). Let $\Omega _{\ast }=1.6\quad \left(
X_{\ast }=2.56\right) $. We get from Eqns. (\ref{sol2a}), (\ref{sol2b}) for $%
X=X_{\ast }$\ one physical solution: $U_{\ast }=-2.\,\allowbreak
946\,246\,654$, $h_{\ast }=0.894\,933\,811\,3$, $\gamma _{\ast
}=2.\,\allowbreak 235\,385\,759$, $Y_{\ast }=0.818\,337\,794\,$\ $\left(
A_{\ast }=0.904620248\right) $.

In Fig. 1 we plot amplitude profiles, i.e. variables $A,\ \Omega $
fulfilling (\ref{AmplDuff}), for the critical value $\gamma =\gamma _{\ast }$
and $h=0.4,\ 0.8,\ 0.85$ and the critical value $h=h_{\ast }$.

\begin{figure}[ht!]
\begin{equation*}
\includegraphics[width= 8 cm, height= 6 cm]{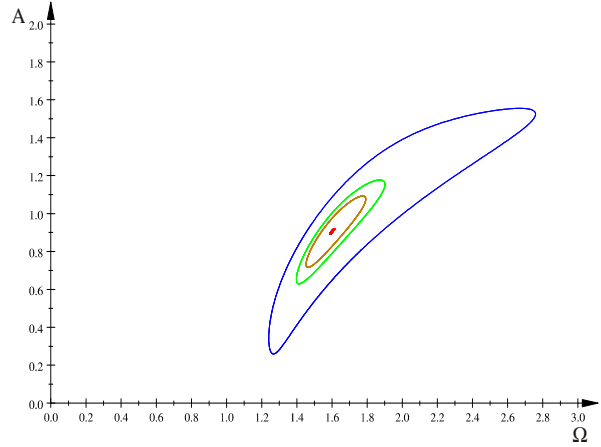}
\end{equation*}%
\caption{Amplitude profiles $A\left( \Omega \right) $, $\protect\gamma =%
\protect\gamma _{\ast }$, $h=0.4$ (blue), $0.8$ (green), $0.85$ (sienna), $%
h_{\ast }$ (red).}
\label{F1}
\end{figure}

\newpage

\begin{figure}[ht!]
\begin{equation*}
\includegraphics[width= 8 cm, height= 6 cm]{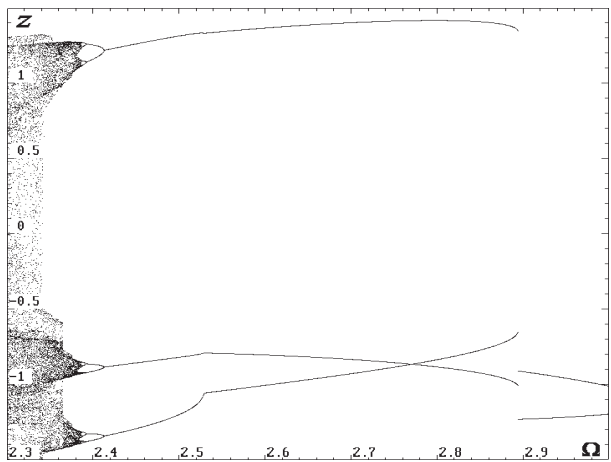}
\end{equation*}%
\caption{Bifurcation diagram for the Duffing equation, $h=0.4,\ \protect%
\gamma =\protect\gamma _{\ast }$.}
\label{F2}
\end{figure}

Resonance $1:3$ is shown in Fig. 2 where bifurcation diagram for the Duffing
equation (\ref{Duffing}) in $\left( z,\ \Omega \right)$ plane was computed
for $h=0.4,\ \gamma =\gamma _{\ast }$ and $\Omega \in \left[ 2.3,\ 3.0 %
\right] $.

Since the KBM method is approximate metamorphosis in the real system may
happen at a slightly different value of, say, parameter $h$. The numerically
exact critical value of this parameter was determined from the bifurcation
diagram below where dependence of $z$ on $h$ is shown for $\Omega =\Omega
_{\ast }$, $\gamma =\gamma _{\ast }$. \bigskip

\begin{figure}[ht!]
\begin{equation*}
\includegraphics[width= 8 cm, height= 6 cm]{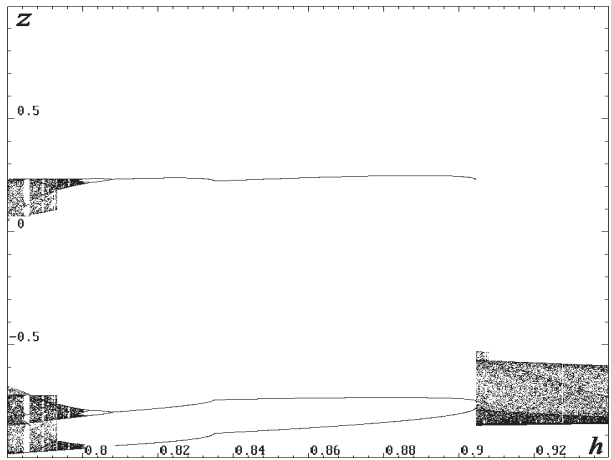}
\end{equation*}%
\caption{Bifurcation diagram for the Duffing equation, $\Omega = \Omega
_{\ast }$, $\protect\gamma=\protect\gamma _{\ast }$.}
\label{F3}
\end{figure}

It follows that the $1:3$ resonance ends abruptly for growing $h$ at $h\cong
0.905$, i.e. slightly above the critical value $h_{\ast}$.

In Figs. 4, 5 below bifurcations diagrams showing dependence of $z$ on $%
\Omega $ for $\gamma =\gamma _{\ast }$ and $h=0.90,\ 0.92$ are shown.

\begin{figure}[ht!]
\begin{equation*}
\includegraphics[width= 8 cm, height= 6 cm]{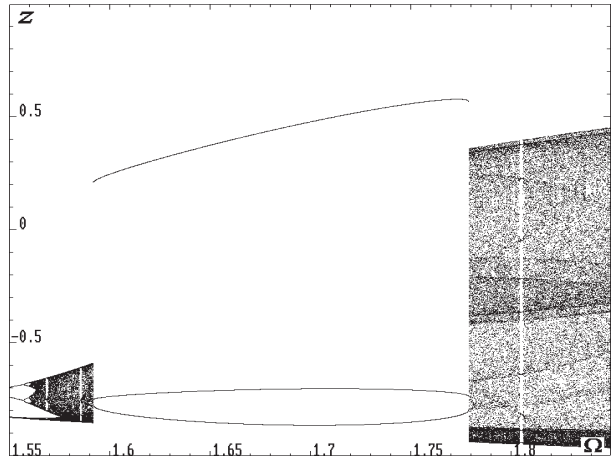}
\end{equation*}%
\caption{Bifurcation diagram for the Duffing equation, $h=0.9,\ \protect%
\gamma =\protect\gamma _{\ast }$.}
\label{F4}
\end{figure}

We realize that the $1:3$ resonance disappears for growing $h$ in agreement
with analytical computations (based however on the approximate KBM method).

\begin{figure}[ht!]
\begin{equation*}
\includegraphics[width= 8 cm, height= 6 cm]{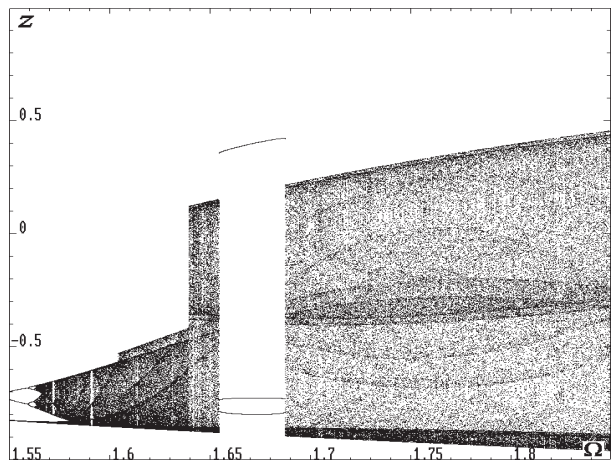}
\end{equation*}%
\caption{Bifurcation diagram for the Duffing equation, $h=0.92,\ \protect%
\gamma =\protect\gamma _{\ast }$.}
\label{F5}
\end{figure}

\section{Summary and discussion}

In this work we have studied metamorphoses of amplitude profiles for the $%
1:3 $ resonances of the effective equation, describing approximately
dynamics of two coupled periodically driven oscillators. Our analysis has
been analytical although based on the approximate KBM method. Theory of
algebraic curves has been used to compute singular points on amplitude
profiles of the effective as well as the Duffing equation. It follows from
general theory that metamorphoses of amplitude profiles occur in
neighbourhoods of such points. The results obtained can be compared with our
work on metamorphoses of $1:1$ resonances in the effective equation \cite%
{Kyziol2011}.

In Section 4 we have computed analytically positions of singular points for
the amplitude profiles $A\left( \Omega \right) $\ determined within the
Krylov-Bogoliubov-Mitropolsky approach for the approximate 2nd-order
effective equation (\ref{effective}). In Section 5 analytical and numerical
results have been presented for the case of the Duffing equation arising as
the subsequent approximation of the effective equation. We have also
computed numerically bifurcation diagrams in the neighbourhoods of singular
points and indeed dynamics of the Duffing equation (\ref{Duffing}) changes
according to metamorphoses of the corresponding amplitude profiles. More
exactly, we have found only the case of isolated singular point and this
corresponds to creation or destruction of $1:3$ resonance. We are going to
investigate much more complicated case of the effective equation in our next
paper.

\end{document}